\documentclass[useAMS,usenatbib]{mn2e}
\usepackage[dvips]{graphicx}
\usepackage{amsmath}
\usepackage{hyperref}
\usepackage{xcolor}
\usepackage{color}
\usepackage{amssymb}
\usepackage{gensymb}
\usepackage{epstopdf}

\newcommand{\bz}{$\langle B_z \rangle$}

\newcommand{\vsini}{$v \sin i$}
\newcommand{\kms}{km\,s$^{-1}$}

\newcommand{\msun}{M$_\odot$}

\newcommand{\teff}{$T_{\rm eff}$}

\newcommand{\mnras}{$MNRAS$}
\newcommand{\apj}{$ApJ$}
\newcommand{\apjl}{$ApJL$}
\newcommand{\aj}{$AJ$}
\newcommand{\aap}{A\&A}

\newcommand{\aaps}{A\&AS}

\newcommand{\pasp}{PASP}

\newcommand{\apss}{ApSS}

\newcommand{\physscr}{Phys.\ Scr.}

\definecolor{alex}{rgb}{0.5, 0.25, 0.25}

\definecolor{matt}{rgb}{1., 0., 0.}

\title[The Bp eclipsing binary HD\,62658]{MOBSTER - III. HD\,62658: a magnetic Bp star in an eclipsing binary with a non-magnetic `identical twin'}
\author[M. Shultz]
{M.\ E.\ Shultz$^{1}$\thanks{E-mail: mshultz@udel.edu}\thanks{Annie Jump Cannon Fellow},
C.\ Johnston$^2$,
J.\ Labadie-Bartz$^3$,
V.\ Petit$^1$
A.\ David-Uraz$^1$, 
\newauthor{O.\ Kochukhov$^4$, G.\ A. Wade$^{5}$, J.\ Pepper$^6$, K.\ G.\ Stassun$^{7,8}$, J.\ E.\ Rodriguez$^9$}
\newauthor{M.\ B.\ Lund$^{10}$, D.\ J. James$^{9,11}$}\\
$^1$Department of Physics and Astronomy, University of Delaware, 217 Sharp Lab, Newark, Delaware, 19716, USA\\
$^2$Instituut voor Sterrenkunde, KU Leuven, Celestijnenlaan 200D, 3001, Leuven, Belgium\\
$^3$Instituto de Astronomia, Geof\'isica e Ciencias Atmosf\'ericas, Universidade de S\`ao Paulo, Rua do Mat\~ao 1226,\\
Cidade Universit\~aria, S\'ao Paulo, SP 05508-900, Brazil\\
$^4$Department of Physics and Astronomy, Uppsala University, Box 516, Uppsala 75120, Sweden \\
$^5$Department of Physics and Space Science, Royal Military College of Canada, Kingston, Ontario K7K 7B4, Canada\\
$^6$Department of Physics, Lehigh University, 16 Memorial Drive East, Bethlehem, PA, 18015, USA\\
$^7$Department of Physics and Astronomy, Vanderbilt University, Nashville, TN 37235, USA\\
$^8$Department of Physics, Fisk University, 1000 17th Avenue North, Nashville, TN 37208, USA\\
$^9$Center for Astrophysics \textbar \ Harvard \& Smithsonian, 60 Garden St, Cambridge, MA 02138, USA\\
$^{10}$Caltech IPAC – NASA Exoplanet Science Institute 1200 E. California Ave, Pasadena, CA 91125, USA\\
$^{11}$Black Hole Initiative at Harvard University, 20 Garden Street, Cambridge, MA 02138, USA\\}
\begin{document}

\date{}

\pagerange{\pageref{firstpage}--\pageref{lastpage}} \pubyear{2019}

\maketitle

\label{firstpage}

\begin{abstract}
HD\,62658 (B9p\,V) is a little-studied chemically peculiar star. Light curves obtained by the Kilodegree Extremely Little Telescope (KELT) and Transiting Exoplanet Survey Satellite (TESS) show clear eclipses with a period of about 4.75~d, as well as out-of-eclipse brightness modulation with the same 4.75~d period, consistent with synchronized rotational modulation of surface chemical spots. High-resolution ESPaDOnS circular spectropolarimetry shows a clear Zeeman signature in the line profile of the primary; there is no indication of a magnetic field in the secondary. PHOEBE modelling of the light curve and radial velocities indicates that the two components have almost identical masses of about 3 \msun. The primary's longitudinal magnetic field $\langle B_z \rangle$ varies between about $+100$ and $-250$ G, suggesting a surface magnetic dipole strength $B_{\rm d} = 850$~G. Bayesian analysis of the Stokes $V$ profiles indicates $B_{\rm d} = 650$~G for the primary and $B_{\rm d} < 110$ G for the secondary. The primary's line profiles are highly variable, consistent with the hypothesis that the out-of-eclipse brightness modulation is a consequence of rotational modulation of that star's chemical spots. We also detect a residual signal in the light curve after removal of the orbital and rotational modulations, which might be pulsational in origin; this could be consistent with the weak line profile variability of the secondary. This system represents an excellent opportunity to examine the consequences of magnetic fields for stellar structure via comparison of two stars that are essentially identical with the exception that one is magnetic. The existence of such a system furthermore suggests that purely environmental explanations for the origin of fossil magnetic fields are incomplete.
\end{abstract}


\section{Introduction}

Surface magnetic fields are detected in about 10\% of main sequence stars with radiative envelopes \citep{2017MNRAS.465.2432G,2019MNRAS.483.2300S}. These magnetic fields are typically strong \citep[above 300~G:][]{2007A&A...475.1053A,2019MNRAS.483.3127S}, globally organized \citep[usually dipolar, e.g.][]{2018MNRAS.475.5144S,2019A&A...621A..47K}, and stable over timescales of at least decades \citep[e.g.][]{2018MNRAS.475.5144S}. These properties, together with the absence of any obvious dependence of surface magnetic field strength upon rotation as would be expected for the dynamo-sustained magnetic fields of cool stars, leads to the characterization of hot star magnetic fields as so-called ``fossil'' fields \citep[e.g.][and references therein]{2015IAUS..305...61N}. 

It is extraordinarily rare to find a magnetic early-type star in a close binary system (i.e.\ and orbital period less than about 1 month). The Binarity and Magnetic Interactions in various classes of Stars (BinaMIcS) survey found an incidence rate of magnetic stars in close binaries below 2\% across the population of upper main sequence multiple systems \citep{2015IAUS..307..330A}. This is a surprising result given that the binary fraction of hot stars is very high \citep[e.g.][]{2012Sci...337..444S,2014ApJ...782....7D}. It has been suggested that this rarity might be related to the formation mechanism for fossil magnetic fields. For instance, if fossil magnetic flux is inherited and amplified from the molecular cloud in which the star is born, strong magnetic fields might inhibit cloud fragmentation and, thus, prevent the formation of close binary systems \citep{2008MNRAS.385.1820P,2011ApJ...742L...9C}. Alternatively, fossil fields might be left over from powerful dynamos generated during stellar mergers, an observation compatible with the apparent anomalous youth of some magnetic stars \citep{2016MNRAS.457.2355S} as well as with the expected rate of mergers \citep{2013ApJ...764..166D,2014ApJ...782....7D}. It has alternatively been suggested that the tidal influence of a close companion might lead to rapid decay of fossil magnetic fields \citep{2019arXiv190210599V}. 

Since only a handful of magnetic close binaries are known (a list is provided by \citealt{2017A&A...601A.129L}), there is value in both increasing the sample of such stars, as well as closely studying the known systems. Recently, examination of the Kilodegree Extremely Little Telescope \citep[KELT;][]{2007PASP..119..923P} light curve of the little-studied star HD\,62658 revealed the presence of eclipses as well as out-of-eclipse brightness modulations. This pattern is very similar to that observed in the first discovered eclipsing binary magnetic Ap star, HD\,66051 \citep{2018MNRAS.478.1749K}. Another probable Ap star, HD\,99458, which has a low-mass eclipsing companion, was recently reported by \cite{2019MNRAS.487.4230S}, although magnetic measurements have not yet been obtained.  Since HD\,62658 is listed as a chemically peculiar Bp star in the \cite{2009A&A...498..961R} {\em Catalogue of Ap, HgMn and Am stars}, we obtained high-resolution spectropolarimetric observations in order to search for the presence of a magnetic field. In the following, we refer to the component producing the light curve's rotational modulation as the primary, and the other component as the secondary\footnote{As demonstrated in \S~\ref{sec:phot}, the component responsible for the out-of-eclipse variability is actually slightly less massive; however, the difference is small enough that we maintain this nomenclature for the sake of clarity.}.


The goal of the study presented here, which is the third of a series of publications by the MOBSTER Collaboration\footnote{Magnetic OB[A] Stars with TESS: probing their Evolutionary and Rotational properties; \cite{2019MNRAS.487..304D}.}, is to provide a first characterisation of HD\,62658. In the following we report the results of our observations, together with the recently obtained Transiting Exoplanet Survey Satellite (TESS) light curve. In \S~\ref{sec:obs} we describe the photometric and spectropolarimetric datasets. The KELT and TESS light curves are analysed, and the orbital parameters determined, in \S~\ref{sec:phot}. Magnetometry and magnetic modelling is presented in \S~\ref{sec:mag}. The implications of our results are explored in \S~\ref{sec:discussion}, and our conclusions are summarized in \S~\ref{sec:conclusions}.

\section{Observations}\label{sec:obs}

\subsection{Photometry}

\subsubsection{KELT}

KELT is a photometric survey comprising two similar telescopes. KELT-North \citep{2007PASP..119..923P} is located at Winer Observatory in Sonoita, Arizona, and KELT-South \citep{2012PASP..124..230P} is situated at the South African Astronomical Observatory in Sutherland, South Africa. Both telescopes have a 42 mm aperture, a 26$^{\circ}$ x 26$^{\circ}$ field of view, and a pixel scale of 23$^{\prime\prime}$. The KELT survey is designed to detect giant exoplanets transiting stars with apparent magnitudes between 8 $\lesssim V \lesssim$ 11, and is well-suited for detecting periodic signals in stellar light curves down to amplitudes of a few mmag \citep{2019arXiv190711666L}. The single passband of the KELT telescopes is roughly equivalent to a broadband $V+R+I$ filter. The normal telescope operations are completely automated and observations are made nightly. HD\,62658 was observed 2730 times with KELT-South between 
2013 May 11 -- 2017 Oct 1 with a median cadence of 31 minutes, covering 337 orbital cycles over the observational baseline.

Part of the KELT strategy for discovering transiting exoplanets involves an algorithm that pre-selects potential exoplanet candidates from reduced light curves for all sources identified in a given field \citep{2018AJ....156..234C}. HD\,62658 was one such source identified in this way. However, the light curve clearly shows eclipses of two different depths, and is thus more likely to be an eclipsing binary. The out-of-eclipse variability apparent in the light curve is inconsistent with ellipsoidal variation, since this is due to a geometrical distortion in the stellar surface that is strongest at periastron, and is therefore generally detectable in eccentric binaries; as the eclipses are separated by close to 0.5 orbital cycles, eccentricity should be low and ellipsoidal variation is not expected. A rotational origin was therefore suspected, prompting further investigation.

\begin{figure*}
   \centering
   \includegraphics[width=2.1\columnwidth]{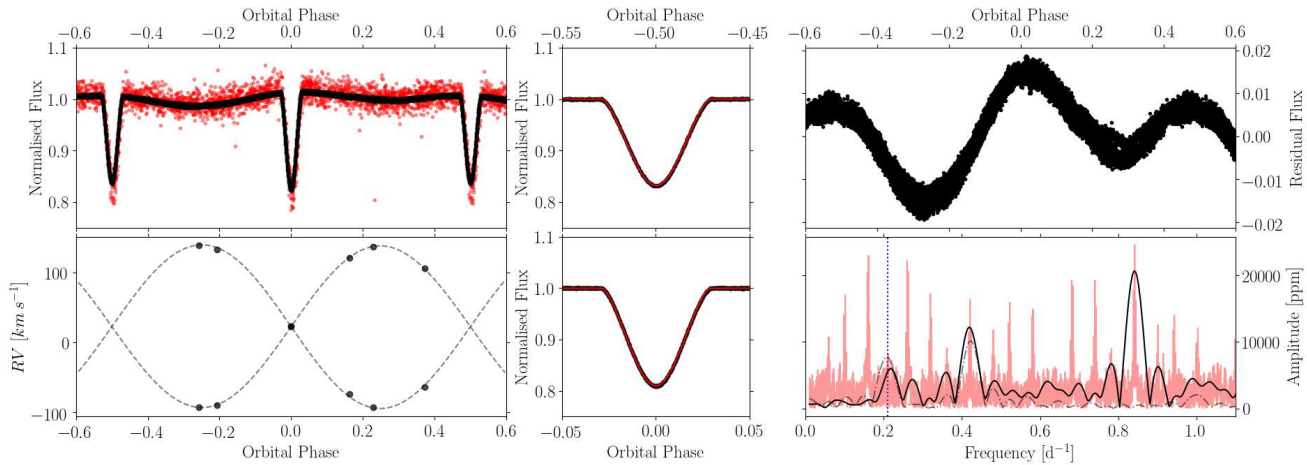}
      \caption[]{{\it Upper left}: Phase folded TESS light curve (black) and KELT light curve (red). {\it Lower left}: Phase folded radial velocity (RV) observations (black circles) with optimized RV curves (dashed black lines). {\it Middle column}: Zoom in of primary (upper panel) and secondary (lower panel) eclipse with optimized model (red). {\it Upper right}: Residual light curve after removal of binary model, phase folded over the orbital period. {\it Lower right}: Periodograms of the original KELT data (red), original TESS data (black), and TESS data after removal of the binary model (dashed gray). The orbital frequency is denoted by vertical blue dashed line. Harmonics of the main period and multiple aliases are apparent, especially associated with the diurnal observing strategy of KELT.}
         \label{hd62658_lc_residuals}
   \end{figure*}

The KELT data are shown phased with the orbital period in the top left panel of Fig.\ \ref{hd62658_lc_residuals}. Using the Python package ${\rm astropy}$ \citep{astropy:2013,astropy:2018}, a Lomb-Scargle \citep{Lomb1976, Zechmeister2009} frequency analysis of the data using a single Fourier term reveals a periodogram with many peaks (bottom right panel of Figure~\ref{hd62658_lc_residuals}), including that associated with the 0.21043(4) d$^{-1}$ orbital frequency. Inspection of the light curve phased to the remaining peaks reveals that they are either harmonics of the orbital period or are aliases induced by the observing strategy of KELT (the most prominent being at 1 and 2 d$^{-1}$). 

\subsubsection{TESS}

Launched on 18 April 2018, TESS seeks to discover new exoplanets by surveying about $\sim$85\% of the sky over its 2-year nominal mission, divided into 26 partially overlapping ``sectors" (each corresponding to a total field of view of 24$\degree\times$96$\degree$ across the four cameras onboard; the pixel size is 21$''$) that are each observed for $\sim$ 27~d \citep{2015JATIS...1a4003R}. The TESS bandpass is broad and covers a range of approximately 6,000-10,000 \AA. Full-frame images (FFIs) are acquired every 30 minutes. Over 500 million point sources fall into at least one of these sectors (and are thus included in the TESS Input Catalog, or TIC), and out of these, $\sim$ 200,000 were selected for 2-minute cadence observations \citep{2018AJ....156..102S}.

HD\,62658 (= TIC 149319411) is one such target, and was observed by TESS in sectors 7 and 8 (7 Jan. 2019 - 28 Feb. 2019; observing programs G011127 and G011060, PI Ricker). Although no contamination ratio is available for this star in the TIC, it is the brightest star by about 2.5 mag in the TESS bandpass within a radius of 3.5$'$ (or about 10 pixels); therefore, the variations seen in the TESS light curve are likely intrinsic to HD\,62658. The observations for this star were downloaded from the Mikulski Archive for Space Telescopes (MAST)\footnote{\url{http://archive.stsci.edu/}} and we used the \texttt{PDCSAP} flux column from the light curve files generated by the TESS Science Processing Operations Center \citep{2016SPIE.9913E..3EJ}. 

Upon an initial investigation of the TESS light curves, we found the sector 7 photometry to be well behaved, while the sector 8 data exhibited strong instrumental trends near the gaps that are present between sectors and in the middle of each sector. Taking into account data points marked as poor-quality in the TESS light curve files, the sector 8 data also feature a significantly larger middle gap (5.9~d) compared to the sector 7 observations (1.7~d). Because of the time scales associated with the different sources of variability described in \S~\ref{sec:phot}, properly detrending these artefacts without affecting the signals we are attempting to model would prove to be quite difficult, and as such, we consider this effort to lie outside the scope of this initial discovery paper. Therefore, we chose to only take into account the data acquired in sector 7. This decision does not severely impact the scientific yield of our study, as the exquisite data quality of the TESS light curve allows us to detect low amplitude signals, while the long temporal baseline of the KELT data can be leveraged to accomplish a very precise orbital period determination. The sector 7 data (black) are shown together with the KELT data (red) in the top panel of Fig.\ \ref{hd62658_lc_residuals}.

\subsection{Spectropolarimetry}

\begin{table*}
\centering
\caption[\bz~measurements]{ESPaDOnS observation log and table of RV and \bz~measurements.`DF' is the detection flag (described in more detail in the text). \bz~measurements were performed using the measured equivalent width of the Stokes $I$ profiles; note that the observation of 21/03 was obtained while the secondary was eclipsing the primary, and the \bz~measurement at this time is therefore not reliable. RV uncertainties are estimated at 0.8 \kms~for the primary and 0.6 \kms~for the secondary, as determined from the standard deviation across 10 fits to the LSD Stokes $I$ profiles.}
\begin{tabular}{l l r r r l r r l}
\hline\hline
        &      &     & \multicolumn{3}{c}{Primary} & \multicolumn{3}{c}{Secondary} \\
HJD -   & Date & S/N & RV     & \bz & DF & RV     & \bz & DF \\ 
2458500 &      &     & (\kms) & (G) &           & (\kms) & (G) &           \\ 
\hline
57.74908 & 15/03/2019 & 236 &  139 & $  37 \pm  51$ &  DD &  -92 & $  31 \pm  83$ &  ND \\ 
59.74646 & 17/03/2019 & 240 & -73 & $ 121 \pm  39$ &  DD &  121 & $  73 \pm  79$ &  ND \\ 
60.74559 & 18/03/2019 & 275 & -63 & $-258 \pm  39$ &  DD &  106 & $ 165 \pm  75$ &  ND \\ 
62.74468 & 20/03/2019 & 293 & 133 & $ 102 \pm  34$ &  DD &  -89 & $-135 \pm  62$ &  ND \\ 
63.72817 & 21/03/2019 & 272 &  22 & $  96 \pm  26$ &  MD &   22 &  -- & -- \\
64.81290 & 22/03/2019 & 298 &  -93 & $  13 \pm  34$ &  DD &  137 & $   2 \pm  58$ &  ND \\    
\hline\hline
\end{tabular}
\label{bzrv}
\end{table*}

Between 15/03/2019 and 22/03/2019, six spectropolarimetric circular polarization (Stokes $V$) sequences of HD\,62658 were obtained with the ESPaDOnS instrument at the Canada-France-Hawaii Telescope (CFHT) under program code 19AC19. The observation log is provided in Table \ref{bzrv}. ESPaDOnS is a high resolution ($\lambda/\Delta\lambda \sim 65000$ at 500 nm) echelle spectropolarimeter covering the spectral range between 370 and 1000 nm across 40 spectral orders. The reduction and analysis of ESPaDOnS data were described in detail by \cite{2016MNRAS.456....2W}. Each observation consists of 4 unpolarized Stokes $I$ spectra, one Stokes $V$ spectrum, and two diagnostic null $N$ spectra obtained by combining the different polarizations in such a way as to cancel out the intrinsic polarization of the source. 

A uniform sub-exposure time of 597 s was used for each sub-exposure, with the total exposure time across the sequence 4$\times$ this number (i.e.\ 2388 s). The mean peak per pixel signal-to-noise (S/N) ratio in the dataset is 269; all 6 observations are of comparable quality (see Table \ref{bzrv}).

Each observation was post-processed by normalizing each spectral order using polynomial splines fit by eye to the continuum, thus ensuring the continuum is as close as possible to unity, while avoiding as much as possible over-normalization due to broad features such as H Balmer wings at the edges of spectral orders. 

\section{Light curve analysis}\label{sec:phot}

   \begin{table}
    \centering
    \renewcommand{\arraystretch}{1.3}
        \caption{Parameters varied during the MCMC optimisation.
             All parameters correspond to median values, with errors listed as the
             boundaries taken from 68.27\% HPD intervals.}

    \begin{tabular}{l c c l}
    \hline\hline
    Parameter &  & Prior Range & HPD Estimate \\    
    \hline
    $T_0-2456425$ & d & (-2,2) & $0.231\substack{+0.005 \\ -0.006}$ \\
    $P_{\rm orb}$ & d & (4,6) & $4.752212\substack{+1e-5 \\ -9e-6}$ \\
    $q$ & $\frac{M_2}{M_1}$ & (0.5,1.5) & $1.012\substack{+0.006 \\ -0.007}$ \\
    $a$ & $R_{\odot}$ & (10,30) & $22.041\substack{+0.001 \\ -0.001}$ \\
    $\gamma$ & ${\rm km\,s^{-1}}$ & (-50,50) & $22.8\substack{+0.4 \\ -0.4}$ \\
    $i$ & $\deg$ & (70,90) & $83.6027\substack{+0.01 \\ -0.008}$ \\
    $e\cos\omega_0$ & & (-0.1,0.1) & $0.00022\substack{+1e-5\\-1e-5}$\\
    $e\sin\omega_0$ & & (-0.1,0.1) & $0.0042\substack{+0.0004\\-0.0004}$\\
    $T_{\rm eff,1}$ & ${\rm K}$ & N/A & $12\,500$ \\
    $T_{\rm eff,2}/T_{\rm eff,1}$ &  & (0.5,1.5) & $0.9385\substack{+0.005 \\ -0.005}$ \\
    $\Omega_1$ &  &  (5,20) & $10.58\substack{+0.01\\-0.01}$\\
    $\Omega_2$ &  &  (5,20) & $10.39\substack{+0.06\\-0.05}$\\
    $\omega_{\rm rot,1}/\omega_{\rm orb}$ & & (0.2,5) & $1.07\substack{+0.05\\-0.06}$ \\
    $\omega_{\rm rot,2}/\omega_{\rm orb}$ & & (0.2,5) & $0.81\substack{+0.03\\-0.03}$ \\
    $l_1$ & \% & (20,80) & $51.1\substack{+0.2\\-0.2}$\\
    $l_3$ & \% & (0,20) & $2.2\substack{+0.3\\-0.3}$\\
    \hline\hline
    \end{tabular}
    \label{tab:binary_sampled_pars}
\end{table}

\begin{table}
    \centering
    \renewcommand{\arraystretch}{1.3}
        \caption{Geometric and derived parameters and their values
            as obtained from MCMC modelling.}
    \begin{tabular}{l c c}
    \hline\hline
    Parameter & & Estimate \\    
    \hline
    $e$& & $0.0042\substack{+0.0004\\-0.0004}$\\
    $\omega_0$& ${\rm rad}$ & $1.519\substack{+0.005\\-0.005}$\\
    $r_2/r_1$ & & $1.031\substack{+0.002\\-0.002}$\\
    $r_1+r_2$ & & $0.2126\substack{+0.0001\\-0.0001}$\\
    $M_1$ & ${\rm M_{\odot}}$ & $3.16\substack{+0.01\\-0.01}$\\
    $M_2$ & ${\rm M_{\odot}}$ & $3.20\substack{+0.01\\-0.01}$\\
    $R_1$ & ${\rm R_{\odot}}$ & $2.307\substack{+0.002\\-0.003}$\\
    $R_2$ & ${\rm R_{\odot}}$ & $2.377\substack{+0.003\\-0.003}$\\
    $\log g_1$ & dex & $4.212\substack{+0.002\\-0.002}$\\
    $\log g_2$ & dex & $4.191\substack{+0.002\\-0.002}$\\
    \hline\hline
    \end{tabular}
    \label{tab:binary_derived_pars}
\end{table}

\begin{table}
    \centering
    \caption{Frequencies, amplitudes, and phases extracted from the TESS light curve with the optimised binary model removed.}
    \begin{tabular}{l c c c c}
    \hline\hline
     & Frequency ${\rm [d^{-1}]}$& Amplitude ${\rm [ppm]}$ & SNR & Note \\    
    \hline

    $f_1$ & $0.21161\pm0.00004$ & $6968\pm96$ & 35 &  $f_{\rm rot}$ \\
    $f_2$ & $0.42324\pm0.00002$ & $9763\pm120$ & 47 & $2f_1$ \\
    $f_3$ & $0.9998\pm0.0001$   & $2073\pm97$ & 15 &  \\

    \hline\hline
    \end{tabular}

    \label{tab:frequency_table}
\end{table}

In order to obtain dynamical mass and radius estimates for the
components of this system, we performed light curve modelling
with the PHOEBE binary modelling code \citep{prsa2005,prsa2011} following the 
framework of \cite{2018MNRAS.478.1749K}, which we briefly 
summarise below. 

\subsection{Signal Separation}
Inspection of the TESS light curve reveals both eclipses and 
apparent spot modulation (see Fig.\ \ref{hd62658_lc_residuals}). Phase folding the light curve on the 
orbital period reveals that the period of the rotational modulation due to spots is nearly commensurate 
with the orbital period. Although PHOEBE can model spots, 
their inclusion in the modelling process can become highly 
degenerate without stringent constraints. As the spectroscopic
dataset cannot provide the 
location, size, temperature, and multiplicity constraints
required by the photometric spot model, we chose to model
the spot signal as a harmonic series instead. 
Since the amplitude of the spot signal in the frequency spectrum is of the same order as 
the orbital signal, we had to disentangle the two iteratively. 
As a first approach, we clipped the eclipses and fit a harmonic
series to the remaining signal via non-linear least squares,
which was then removed from the original light curve. Then, a 
binary model was optimised on these residuals. The binary model
was then removed from the original light curve, and we fit a harmonic
series to these residuals. A new binary model was optimised and the 
process was repeated until there was no change in the resulting 
fit. Since we removed an aphysical harmonic series from 
the light curve, we fixed the albedos and gravity brightening
exponents of the components to unity, i.e.\ we assumed that the out-of-eclipse variability is not due to ellipsoidal variations. 

\subsection{Modelling setup}
To optimise our solution, the PHOEBE binary modelling
code was wrapped into the Bayesian sampling code {\sc emcee}
\citep{foremanMackey2013} which employs an affine-invariant 
Markov Chain Monte Carlo (MCMC) ensemble sampling approach 
to numerically evaluate 
the posterior distribution of a set of sampled parameters. 
The posterior distribution of a set of sampled parameters, 
$p\left(\Theta|d\right)$, is given by Bayes' Theorem:
\begin{equation}
p\left(\Theta|d\right)\propto\mathcal{L}\left(d|\Theta\right)\,p\left(\Theta\right), 
\end{equation}
\label{bayes}

\noindent where $\Theta$ is the vector of sampled parameters which describe the light curve and $d$ are the TESS data. We take the likelihood function $\mathcal{L}\left(d|\Theta\right)$ to be a $\chi^2$ statistic and encode any previously known  information in the priors, $p\left(\Theta\right)$. The light  curve and radial velocity curves were optimised simultaneously (radial velocity measurements were obtained from ESPaDOnS data; see \S~\ref{sec:mag}). As mentioned previously, we fixed the albedos and gravity brightening exponents. We fixed the primary effective temperature to $12\,500$ K and sampled the ratio of the temperatures $T_{\rm eff,2}/T_{\rm eff,1}$. Furthermore, to incorporate as much information as possible, we applied Gaussian priors on the projected rotational velocities $v_{1,2} \sin i$ according to those values derived in \S~\ref{sec:mag}. Finally, we allowed for an eccentric orbit. 

\subsection{Modelling results}

The sampled parameters, their priors, and their posterior estimates
are listed in Table~\ref{tab:binary_sampled_pars}. We also report 
geometric and derived parameter estimates and their errors in 
Table~\ref{tab:binary_derived_pars}. The parameter 
estimates were calculated as the median of the posterior distribution, 
while the uncertainties were calculated as 68.27\% Highest Posterior Density (HPD) 
intervals from the marginalised posterior distribution of a given 
parameter. In the case of normally distributed posteriors, HPD 
estimates will agree with the mean and $1\sigma$ of a Gaussian 
fit to the distribution. In the event of non-normally distributed 
posteriors, however, HPD estimates have the advantage of being 
flexible and being able to capture the breadth of the possible 
solution space, and are capable of producing asymmetric uncertainties.
The marginalised posterior distributions are illustrated in the appendix.

The residuals of the best fit model and the original light curve are
shown in black in the upper right panel of Fig.~\ref{hd62658_lc_residuals}. 
The bottom right panel shows the Scargle periodograms of the original KELT (red), original TESS (black), and residual TESS (dashed-gray) light curves, with the orbital frequency marked with a vertical dashed blue line. The extracted frequencies listed in Table~\ref{tab:frequency_table} were pre-whitened from the light curve according to \cite{degroote2009}. 


\begin{figure}
   \centering
   \includegraphics[width=0.5\textwidth]{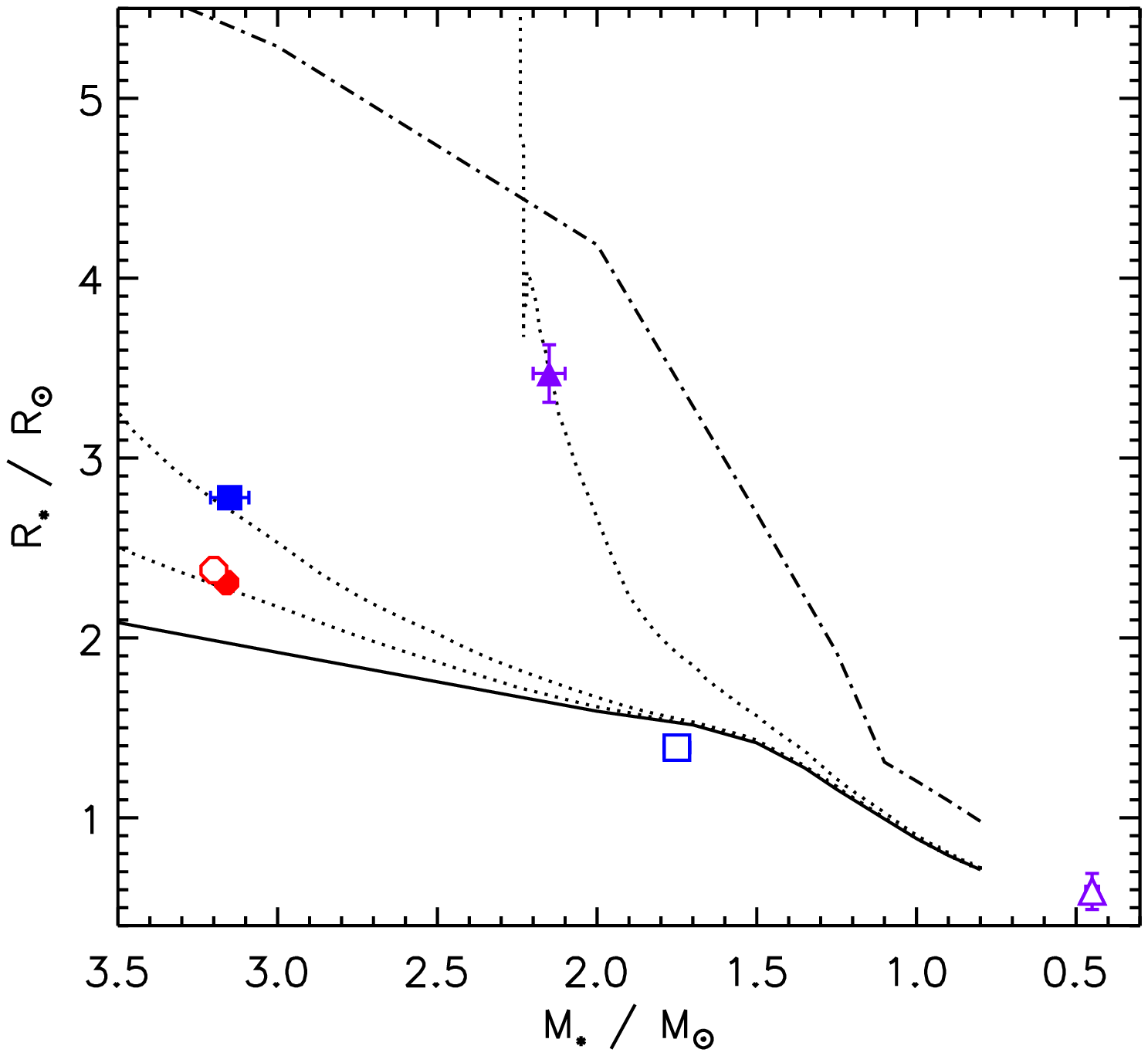}
      \caption[]{Mass-radius diagram of Ap/Bp stars in eclipsing binaries. The ZAMS and TAMS are shown by solid and dot-dashed lines; rotating evolutionary model isochrones \protect\citep{ekstrom2012} by dotted lines, for $\log{(t / {\rm yr})}=8$, 8.3, and 9. HD\,62658 is indicated by red circles, HD\,66051 \protect\citep{2018MNRAS.478.1749K} by blue squares, and HD\,99458 \protect\citep{2019MNRAS.487.4230S} by purple triangles; magnetic and non-magnetic components are indicated by filled and open symbols respectively (although since HD\,99458's secondary is an M dwarf, it presumably hosts a dynamo field, while the magnetic field of the primary is assumed based on its identification as an Ap star).}
         \label{hd62658_mass_radius}
   \end{figure}

We note that $f_1$ and $f_2$ are part of a harmonic series, while $f_3$ 
is an independent signal. Within formal uncertainties, the base frequency 
$f_1$ of the signal attributed to spots is not the same as the orbital 
frequency $f_{\rm orb}=0.2104283(4)$ as determined by {\sc phoebe}. Furthermore, the HPD estimates for 
the synchronicity parameter $\omega_{\rm rot}/\omega_{\rm orb}$ place the 
primary as $1.16\sigma$ away from rotating synchronously. Thus, it is 
reasonable to conclude that the primary is nearly if not already rotating 
pseudo-synchronously with the orbit, within the errors. We note that the 
secondary, however, is (according the the synchronicity parameter) rotating 
sub-synchronously, with a significance of about 6$\sigma$. The derived binary parameters result in the following 
synchronisation and circularisation timescales: $\log(\tau_{\rm sync}/{\rm yr})=6.708\pm0.001$ 
and $\log(\tau_{\rm circ} / {\rm yr})=9.108\pm0.001$ \citep{zahn1975,zahn1977}. As shown in 
Fig. \ref{hd62658_mass_radius}, HD~62658 matches well with a $\log{(t/{\rm yt})}=8$ isochrone, 
suggesting again that the system should be pseudo-synchronized, but not yet 
circularized, as is evidenced by the small eccentricity we find from binary 
modelling. 

As we removed a harmonic series representing the spot signal, and hence the 
non-baseline light, the estimates of third light (i.e.\ the amount of light contributed by a hypothetical third star) are to be considered with 
caution. The independent frequency $f_3$ occurs in the frequency region where 
gravity mode pulsations are expected in slowly pulsating B-type stars, however, due to the 
uncertain amount of third light, we cannot say for certain that these 
signals originate from an identified component of HD~62658.

Removal of a harmonic series also makes direct modelling of any ellipsoidal variation impossible. However, the {\em a posteriori} prediction of the PHOEBE model is that any such variation should have an amplitude of no more than 0.2\% of the normalized flux, i.e.\ an order of magnitude less than the observed out-of-eclipse variation. The assumed absence of this variation should therefore have negligible impact on these results.

We note that any structure in the residuals is likely due to the 
asymmetric blocking and subsequent modulation of light variations from
the surface features (spots) and/or the pulsational signal, should this 
signal originate from a component of this system. One possible means of
accounting for this would be to incorporate Gaussian Processes into the
modelling procedure, however this is beyond the scope of this discovery 
paper. 

\section{Magnetometry}\label{sec:mag}

   \begin{figure*}
   \centering
   \includegraphics[trim = 50 0 50 0, width=0.9\textwidth]{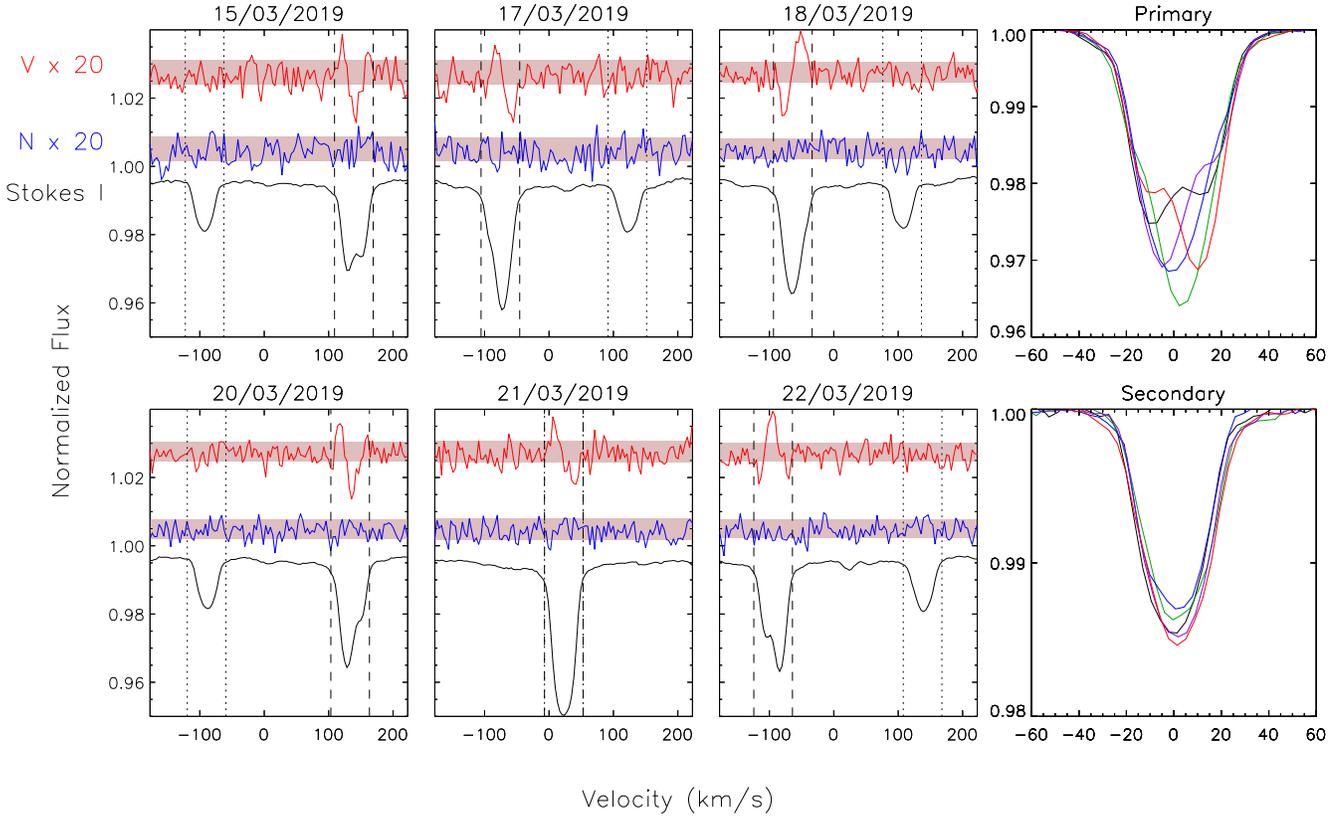}
      \caption[]{LSD profiles extracted from ESPaDOnS spectra. Shaded regions indicate the mean uncertainty. Vertical dashed lines indicate the integration limits of the primary's line profile, dotted lines show the same for the secondary. The two panels on the right show close-ups of the Stokes $I$ LSD profiles of the two components at non-eclipsing phases, shifted to their respective rest frames.}
         \label{hd62658_lsd}
   \end{figure*}

In order to maximize the precision with which the stars' magnetic fields can be measured, least-squares deconvolution \citep[LSD;][]{d1997} profiles were extracted from the ESPaDOnS spectra using the iLSD package \citep{koch2010}. The line mask was created from a line list downloaded from the Vienna Atomic Line Database \citep[VALD3;][]{piskunov1995,ryabchikova1997,kupka1999,kupka2000,2015PhyS...90e4005R} with an `extract stellar' request. We adopted $\log{g}=4.2$ as inferred from the {\sc phoebe} model, and \teff~$=12.3$~kK determined by \cite{1994BSAO...38..152G}, which is consistent with the spectral type of B9p assigned by \cite{2009A&A...498..961R} in their {\em Catalogue of Ap, HgMn and Am stars}. We also adopted enhanced Si, Ti, Cr, and Fe abundances, respectively $[{\rm X/H]=}$ -3.0, -6.0, -5.0, and -3.5, following the \teff-dependent relations for the mean surface abundances of Ap/Bp stars found by \cite{2019MNRAS.483.2300S}. The line depth threshold of the line list is 0.1 below the continuum, as the inclusion of lines weaker than this does not in practice greatly improve the S/N of the LSD profiles, whilst at the same time unreasonably increasing the time taken to extract each profile. The line mask was cleaned using the method described by \cite{2018MNRAS.475.5144S}, with 1581 lines remaining out of the original 2479. LSD profiles were extracted using velocity pixels of 3.6~\kms, or twice the average width of pixels in the extracted ESPaDOnS spectra, in order to slightly decrease the point-to-point scatter, and a Tikhonov regularization factor of 0.2 was applied in order to reduce the signal degradation associated with cross-correlation \citep{koch2010}. 

The resulting LSD profiles are shown in Fig.\ \ref{hd62658_lsd}. The line profiles of the two stellar components are clearly separated in velocity space in five of the observations, and exhibit a radial velocity variation of about $\pm$100 \kms. In one observation (21/03) the line profiles are blended, indicating it was obtained during an eclipse.  Radial velocity (RV) measurements obtained from the LSD Stokes $I$ profiles are given in Table \ref{bzrv}. RVs were measured using the parameterized line profile fitting package described by \cite{2017MNRAS.465.2432G}, which also provides the projected rotational velocities \vsini: for the primary, $26.2 \pm 1.3$~\kms, and for the secondary, $20.4 \pm 0.7$~\kms, where the uncertainties correspond to the standard deviation of the fits across the 5 non-eclipsing observations. The line profile variability of the primary introduces an additional source of systematic uncertainty into \vsini, such that the standard deviation may not fully account for the total uncertainty, although it is worth noting that its uncertainty is almost twice that of the secondary's and therefore the additional uncertainty might already be accounted for. The slightly lower \vsini~of the secondary could be consistent with sub-synchronous rotation of this component. Since macroturbulence is not expected in late B-type stars, this parameter was fixed to 1 \kms~in the profile fitting. Relaxing this constraint increases the uncertainties in \vsini, and leads to macroturbulent velocities of $18 \pm 8$~\kms and $17 \pm 2$~\kms~for the primary and secondary, respectively (which are very high for such stars, and probably unreliable since they were obtained from LSD profiles). 


A Zeeman signature is clearly visible in Stokes $V$ in all observations, corresponding to the position of the primary's line profile. Five observations yield a statistical definite detection (DD) inside the primary's line profile, according to the criteria described by \cite{1992AA...265..669D,d1997} (i.e.\ a False Alarm Probability FAP~$<10^{-5}$). Within the secondary's line profile there is no indication of a Zeeman signature, and these observations yield formal non-detections (NDs) according to the same criteria (FAP~$>10^{-3}$). This indicates that only the primary is detectably magnetic. Detection flags are given in Table \ref{bzrv}.

The blended observation on 21/03 yields a marginal detection (MD). This observation was obtained when the presumed non-magnetic star was eclipsing the magnetic star, and the MD is almost certainly due to light from the magnetic component. As is clear from the light curve, the system is not fully eclipsing (the reduction in flux is only about 15\%), so light from the eclipsed component during eclipses is expected.

The right panels of Fig.\ \ref{hd62658_lsd} show the Stokes $I$ line profiles of the two components, from the five non-eclipsing observations, shifted to their respective rest frames. The complex structure and asymmetry in the primary's line profiles is consistent with the presence of chemical spots. This is consistent with the dominant out-of-eclipse variation in the system's light curve being a consequence of rotational modulation of the primary's surface chemical abundance inhomogeneities. The secondary, by contrast, shows some signs of variability near the core of the line, albeit much weaker than the variations of the primary. These could be a consequence of possible non-radial pulsation identified in the light curve ($f_3$, Table \ref{tab:frequency_table}). 

\begin{figure}
   \centering
   \includegraphics[width=0.5\textwidth]{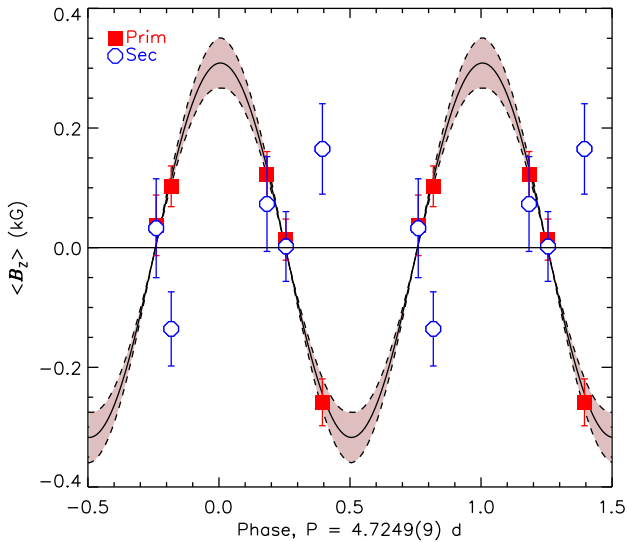}
      \caption[]{Out-of-eclipse \bz~measurements for the two components, phased with the orbital period. The solid curve shows the best-fit sinusoid; the dashed curves show the 1$\sigma$ uncertainties in the fit.}
         \label{hd62658_bz}
   \end{figure}
   
\subsection{Longitudinal magnetic field}

To quantify the strength of the stars' magnetic fields we measured the disk-averaged longitudinal magnetic fields \bz~\citep{mat1989}. These are summarized in Table \ref{bzrv}. Due to the larger uncertainties, the secondary's \bz~is consistent with zero. The primary varies between about \bz~$=+100$ and $-200$~G, and 2 of the observations yield \bz~close to zero (with crossover signatures detectable in Stokes $V$). This indicates we are seeing both magnetic poles and the magnetic equator. The \bz~measurement from the final observation was assigned to the primary, but is not particularly meaningful since the magnetic component is partially obscured.

Fig.\ \ref{hd62658_bz} shows the \bz~measurements of the two components phased with the rotational frequency $f_1$ identified from the light curve (see Table \ref{tab:frequency_table}), corresponding to a period of 4.7249(9)~d. The measurements were phased using $T_{\rm 0, mag} = 2458558.9(2)$, defined at \bz~$=$~\bz$_{\rm max}$ as inferred from a sinusoidal fit. The \bz~measurements of the primary vary coherently with this period. The small difference between $f_1$ and $f_{\rm orb}$ makes essentially no difference in the phasing of \bz, i.e.\ it is not possible using \bz~to test the hypothesis that $f_1 = f_{\rm orb}$.

Fitting a first-order sinusoid of the form \bz~$=B_0 + B_1 \sin{(\phi + \Phi)}$, where $\phi$ is the rotational phase and $\Phi$ is a phase offset, yields $B_0 = 6 \pm 9$~G and $B_1 = 316 \pm 21$~G. The $r$ parameter, used to constrain the relationship between $i_{\rm rot}$ and the magnetic obliquity angle $\beta$ \citep{preston1967},

\begin{equation}\label{preston_r}
r = \frac{|B_0| - B_1}{|B_0| + B_1} = \frac{\cos{(i_{\rm rot} + \beta)}}{\cos{(i_{\rm rot} - \beta)}},
\end{equation}

\noindent is then $r=-0.96 \pm 0.06$. To determine the star's oblique rotator model parameters, we utilized the Hertzsprung-Russell Monte Carlo sampler described by \cite{2019arXiv190902530S}, which provides fully self-consistent magnetic, rotational, and stellar parameters via simultaneous inclusion of all available observables, interpolation through evolutionary models, and probabilistic rejection of inconsistent points in phase space. We adopted the radius from the {\sc phoebe} orbital model, with the luminosity inferred from $R_*$ and \teff, and \vsini~as determined from the fits to the LSD Stokes $I$ profiles. This yielded $i_{\rm rot}=79^\circ \pm 6^\circ$, which is within 1$\sigma$ of $i_{\rm orb}$, consistent with the spin and orbital axes being aligned. The obliquity angle of the magnetic axis from the rotational axis is $\beta = 86^{\circ+14}_{-22}$. The surface polar strength of the magnetic dipole is $B_{\rm d} = 880^{+780}_{-160}$~G, calculated using the observed \bz$_{\rm max} = -258 \pm 39$~G and the linear limb darkening coefficient $\epsilon = 0.47$ from the $B$ band tables calculated by  \cite{diazcordoves1995} (where the $B$ band approximately corresponds to the ESPaDOnS wavelength region containing the majority of spectral lines). Since $i_{\rm rot}$ is consistent with $i_{\rm orb}$, it is reasonable to expect that spin and orbital axes might be exactly aligned. If we take $i_{\rm rot} = i_{\rm orb}$, we find $\beta = 79^{\circ+17}_{-14}$ and $B_{\rm d} = 880 \pm 170$~G. 



To place upper limits on $B_{\rm d}$ for the non-magnetic star, we assumed that $i_{\rm rot}$ is within $10^\circ$ of $i_{\rm orb}$, and adopted the same value of $\epsilon$ as for the magnetic star. The assumption that $i_{\rm rot} \sim i_{\rm orb}$ is justified given that 1) \vsini~differs by only a few \kms~between the two components, and 2) the value of $\omega_{\rm rot}/\omega_{\rm orb}$ inferred from {\sc phoebe} modelling is very close to 1. $B_0=10 \pm 30$~G and $B_1 = 100 \pm 40$~G were taken respectively to be the weighted mean and weighted standard deviation of \bz, with \bz$_{\rm max}$ set to the same value as $B_1$. This yielded 1 and 3$\sigma$ upper limits on $B_{\rm d}$ of 700 G and 1500 G, similar to the value inferred for the magnetic star. Therefore, on the basis of \bz~alone it cannot be ruled out that the non-magnetic star has a magnetic field approximately as strong as that of the magnetic star. 

\subsection{Bayesian modelling of line profiles}

\begin{figure}
   \centering
   \includegraphics[width=0.45\textwidth]{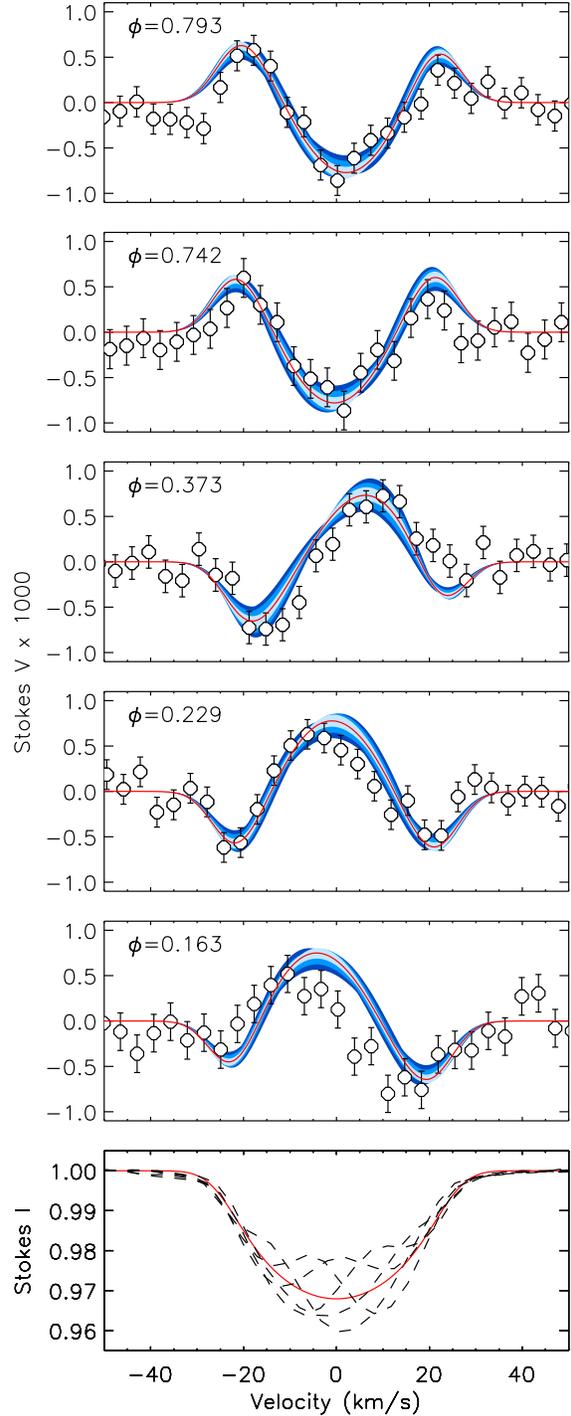}
      \caption[]{Fits to Stokes $V$ from Bayesian modelling. Stokes $I$ is shown in the bottom panel (observed: dashed black; synthetic: solid red). The model does not incorporate line profile variation due to chemical spots. The top five panels show Stokes $V$, with the rotation phase $\phi$ indicated in the top left corner. Observed Stokes $V$ is shown by open circles. The best-fit model is indicated by solid red lines; 68.3\%, 95.4\%, and 99.7\% uncertainties are indicated by light blue, blue, and dark blue shaded regions.}
         \label{hd62658_lsd_bayes_fit}
   \end{figure}
   
As a more precise means of constraining the surface magnetic fields of the two stars, we modelled their Stokes $V$ profiles using the Bayesian inference method described by \cite{petit2012a}. We adopted the same \vsini~values and limb darkening as determined above. Synthetic profile equivalent widths were normalized to the mean value of the dataset. The observation obtained on 21/03/2019 was excluded as the secondary was eclipsing the primary at this time.

For the secondary, this analysis finds 1, 2, and 3$\sigma$ upper limits on $B_{\rm d}$ of 110 G, 340 G, and 1260 G. This field is thus almost certainly below the 300 G critical field limit identified by the survey of weak-field Ap/Bp stars conducted by \cite{2007A&A...475.1053A},  and verified by the volume limited survey of Ap stars presented by \cite{2019MNRAS.483.3127S}. As such, if the star has a magnetic field it is likely to be of the ultra-weak variety exhibited by Vega, Sirius, or Alhena, i.e.\ on the order of 0.1 to 10 G \citep{2010AA...523A..41P,2011AA...532L..13P,2016MNRAS.459L..81B}. 

For the primary we initially performed a fit without constraints on $P_{\rm rot}$ or $i_{\rm rot}$, obtaining maximum-likelihood values for $i$, $\beta$, and $B_{\rm d}$ of about 90$^\circ$, 75$^\circ$, and 500 G. The method therefore strongly prefers a large $i_{\rm rot}$ and $\beta$, with the former consistent with expected spin-orbit alignment. 

In an effort to improve the constraints, we next utilized a modified version of the Bayesian inference code that includes rotational phase information. We also fixed $i_{\rm rot} = i_{\rm orb} = 84^\circ \pm 1$. The resulting fit to Stokes $V$ is shown in Fig.\ \ref{hd62658_lsd_bayes_fit}. This yielded $\beta = 94^\circ$, with 68.3\%, 95.4\%, and 99.7\% uncertainties of $23^\circ$, $46^\circ$, and $60^\circ$. For $B_{\rm d}$ we obtained a maximum posterior probability of 650~G, with upper uncertainties of 150 G, 400 G, and 1000 G, and lower uncertainties of 50 G, 100 G, and 150 G. 

Our Bayesian analysis yields similar values of $\beta$ and $B_{\rm d}$ to those inferred from modelling \bz, with the two overlapping at the 1$\sigma$ level. In contrast to the constraints from \bz, direct modelling of Stokes $V$ is able to demonstrate that any magnetic field present in the atmosphere of the secondary is much weaker than that of the primary, with the difference significant at the 2$\sigma$ level; this is because there are magnetic configurations that yield \bz~$=0$, but still give a detectable Stokes $V$ signal.






\section{Discussion}\label{sec:discussion}

\begin{figure*}
   \centering
   \includegraphics[width=\textwidth]{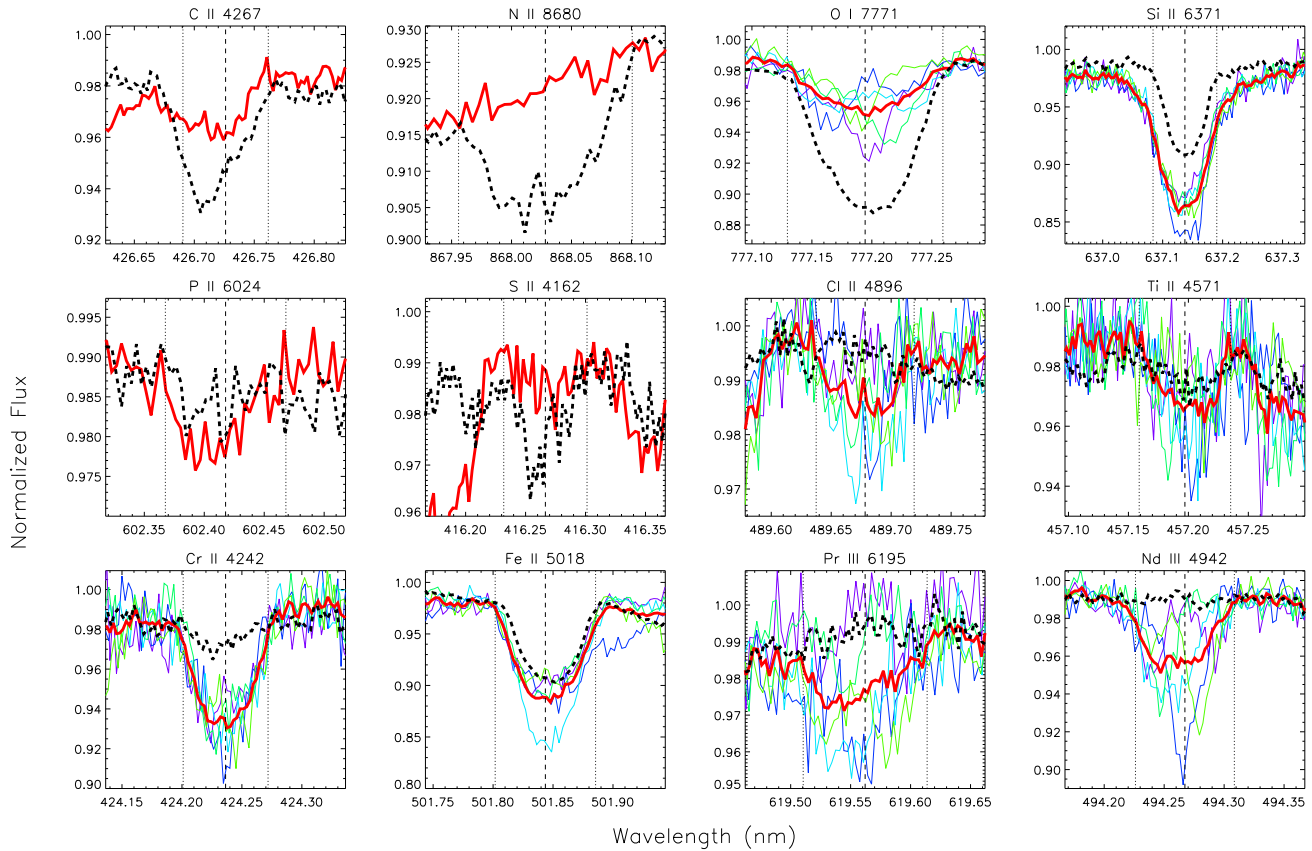}
      \caption[]{Comparison of various spectral lines, obtained from the mean co-added spectrum. Thick red shows the the magnetic primary, thick dashed black lines the non-magnetic secondary. Vertical dashed lines indicate line centre, and dotted lines show the expected line width. In some panels, coloured lines show individual observations shifted to the rest frame of the primary.}
         \label{hd62658_line_compare}
   \end{figure*}

While the variability of the LSD Stokes $I$ profiles (Fig.\ \ref{hd62658_lsd}) is consistent with the magnetic star being a typical Bp star, it is of interest to evaluate whether it demonstrates the typical pattern of chemical abundances seen in such stars. Fig.\ \ref{hd62658_line_compare} compares the line profiles of the magnetic and non-magnetic components. For each component a mean spectrum was created by co-adding the five spectra obtained at non-eclipsing orbital phases, with each spectrum shifted to the laboratory rest frame. Lines were chosen so as to be relatively strong and, more importantly, isolated, with the criterion that there be no strong lines in the VALD line mask within 0.1 nm of the line in question. The non-magnetic star has stronger C~{\sc ii} and O~{\sc i} lines. N~{\sc ii} and S~{\sc ii} are apparently entirely absent in the magnetic star's spectrum. P~{\sc ii}, Ti~{\sc ii}, and Fe~{\sc ii} are similar between the two stars, or slightly stronger in the magnetic star. Si~{\sc ii}, Cl~{\sc ii}, and Cr~{\sc ii} are all stronger or much stronger in the magnetic star. Finally, the magnetic star displays prominent rare earth elements (Pr~{\sc iii} and Nd~{\sc iii}), which are entirely absent in the non-magnetic star.

In some panels of Fig.\ \ref{hd62658_line_compare}, individual spectra shifted to the rest frame of the magnetic star are shown. These demonstrate that O, Si, Fe, Pr, and Nd are all variable, with the variability of O, Fe, Pr, and Nd being particularly strong. Variability is difficult to distinguish from noise in the cases of Cl, Ti, and Cr. This indicates that the elements are not homogeneously distributed across the stellar surface. The variety of line profile morphologies further indicates that the chemical abundance patches are not all distributed uniformly across the stellar surface. A similar evaluation of the line profile variability of the non-magnetic star did not reveal anything obviously different from noise, as expected given the low level of variation in the star's LSD Stokes $I$ profiles.

Fig.\ \ref{hd62658_mass_radius} compares the derived stellar parameters of HD\,62658 to the other Ap/Bp eclipsing binaries, HD\,66051 and HD\,99458, and to isochrones calculated with Geneva evolutionary models \citep{ekstrom2012}. The HD\,62658 components have the same mass as the magnetic component of HD\,66051, but are somewhat younger ($\log{(t / {\rm yr})}=8$ vs.\ 8.3). Both systems are much younger than HD\,99458 ($\log{(t / {\rm yr})}=9$). In contrast to HD\,66051, which has a mass ratio of $q = M_2/M_1 = 0.55$, and of course HD\,99458 for which $q=0.21$, the components of HD\,62658 are essentially identical in mass ($q=1.012 \pm 0.007$). Indeed, while there are several magnetic close binary systems with mass ratios much closer to 1 than that of HD\,66051, e.g. HD\,149277 \citep[$q=0.91$;][]{2016PhDT.......390S,2018MNRAS.481L..30G} and $\epsilon$ Lupi \citep[$q=0.83$;][]{2019MNRAS.488...64P}, the mass ratio $q$ of HD\,62658 is closer to unity than any other known magnetic hot binary.

In addition to having essentially identical masses, HD\,62658's components are presumably coeval \citep[e.g.][]{2001ApJ...556..265W}. Their rotational velocities are furthermore almost the same: while the secondary has a slightly lower \vsini~and is probably rotating sub-synchronously, the differences in their rotation seem unlikely to be important from a dynamical perspective. They apparently differ only in that one of the stars has a fossil magnetic field, and the other does not. This remarkable system may have implications for our understanding of the formation of fossil magnetic fields, and the relation of the mechanism responsible to the overall rarity of magnetic stars in close binary systems. As noted in the introduction, hypotheses seeking to explain the rarity of such systems include:

\begin{enumerate}
    \item Magnetization of the protostellar cloud provides the seed for the fossil field, and also inhibits fragmentation of the cloud and therefore prevents the formation of binaries \citep{2011ApJ...742L...9C}
    \item Fossil fields are left over from dynamos powered by stellar mergers \citep{2016MNRAS.457.2355S}
    \item Tidal interactions in eccentric binaries lead to the rapid decay of fossil magnetic fields \citep{2019arXiv190210599V}
\end{enumerate}

The existence of some, rather than no, close binaries including at least one star with a fossil field suggests that i) is unlikely to be universally true. This depends on whether magnetization prevents fragmentation, or simply makes it unlikely. However, in this scenario, it is curious that one of the components should have inherited all of the pre-stellar magnetic flux, despite the two components being otherwise identical. 

While ii) cannot be ruled out in all cases, it can probably be excluded in magnetic close binaries, as these would need to start as triple (or in the case of the doubly magnetic close binary $\epsilon$ Lupi \citep{2015MNRAS.454L...1S}, quadruple) systems, without the orbit of the merger product(s) being disrupted. The existence of circularized, tidally locked systems containing a magnetic component, such as HD\,66051 and HD\,156324 \citep{2018MNRAS.478.1749K,2018MNRAS.475..839S}, also seems difficult to achieve via mergers.


Since the mechanism suggested in scenario iii) requires variable tidal forces due to an eccentric orbit, it cannot be operating in this case. However, assuming that the system was not always circularized, it may have been operating in the past. The total unsigned magnetic flux of the primary is $\log{\Phi} = \log{[(B_{\rm d}R_*^2) / ({\rm G}~R_\odot^2)]} = 3.6 \pm 0.1$. This is at the lower limit of the range of magnetic fluxes reported by \cite{2019MNRAS.483.3127S} for their volume-limited sample of Ap/Bp stars, which extended up to the mass range occupied by HD\,62658. Since \cite{2019MNRAS.483.3127S} found no evidence for flux decay in this mass range, HD\,62658's magnetic flux is not obviously anomalous for its age. Based on the magnetic and stellar parameters reported by \cite{2018MNRAS.478.1749K}, the unsigned magnetic flux of the slightly older HD\,66051 is about the same as that of HD\,62658, and again at the lower range of the sample presented by \cite{2019MNRAS.483.3127S}. Given the large difference in age between these stars and HD\,99458, it would be of interest to obtain magnetic measurements of the latter.


The previously listed hypotheses for the origin of fossil fields, and their rarity amongst binaries, can broadly be classified as environmental (i.e.\ relating to the magnetic flux within the molecular cloud from which the star formed), or evolutionary (i.e.\ relating to some circumstance of the star's evolution after formation). The existence of a system with stars that are identical in their fundamental parameters, and which must have formed in the same place and at the same time, calls these scenarios into question.

At the bottom of the main sequence there is a magnetic dichotomy somewhat similar to that of hot magnetic stars with and without fossil fields, namely the bimodal distribution of M-dwarf magnetic field strengths and geometries \citep[e.g.][]{2010MNRAS.407.2269M,2011sf2a.conf..503M,2017NatAs...1E.184S}. Some of these stars possess strongly organized poloidal fields with surface strengths above the 4 kG saturation limit, while others have tangled topologies with surface strengths below this limit. This bimodal distribution is thought to be a consequence of a dynamo bistability explored by \cite{2013A&A...549L...5G}, who found that the rotational-convective dynamos of these stars could stabilize into one or the other topology. 

Dynamo bistability is further strengthened by Zeeman Doppler Imaging maps of the M-dwarf binary BL Cet and UV Cet presented by \cite{2017ApJ...835L...4K}. These stars are nearly identical in mass and rotation, yet one possesses a globally organized, axisymmetric poloidal field, while the other has a much weaker, non-axisymmetric, tangled field. The BL Cet/UV Cet system is thus remarkably similar to the case of HD\,62658. Persistent differences in stellar activity indices \citep{2003ApJ...589..983A} suggest that these different magnetic field structures are not due to the stars having been observed at different points in their magnetic activity cycles \citep[although a sudden change in the previously stable axisymmetric magnetic field of of the M dwarf AD Leo was reported by][suggesting that such a change should not be ruled out in the future in the case of BL/UV Cet]{2018MNRAS.479.4836L}. A similar magnetic bistability has been observed by \cite{2018A&A...613A..60R} in the tidally locked F9-G0 system $\sigma^2$ CrB, indicating that a strong sensitivity of dynamo properties on stellar parameters is not limited to fully convective stars.


While fully convective M-dwarves with rotational-convective dynamos, and B stars with radiative envelopes and fossil magnetic fields, are obviously very different in a myriad of important ways, intermediate-mass stars likely pass through a fully convective phase during their pre-main sequence (PMS) evolution. During this period an intermediate mass star is, from a magnetohydrodynamic perspective, somewhat similar to a main sequence M dwarf. It is therefore reasonable to expect that intermediate mass stars may exhibit a similar dynamo bistability on the PMS. This may provide a natural explanation for the ``magnetic desert'' amongst hot main sequence stars, with 10\% possessing globally organized magnetic fields with a lower limit of about 300 G \citep{2007A&A...475.1053A,2014IAUS..302..338L}, and the majority no fields at all, or ultra-weak fields such as those observed on Vega, Sirius, and Alhena \citep{2009AA...500L..41L,2011AA...532L..13P,2016MNRAS.459L..81B}. In this scenario, the fossil magnetic fields of those stars which fail to organize into dipoles, or for which the surface dipole field strength is too low, rapidly decay due to rotationally induced instabilities \citep[e.g.][]{2007A&A...475.1053A,2013MNRAS.428.2789B}. A bistability scenario avoids invoking environmental factors, which seem to be excluded in the case of HD\,62658 since its components are identical in age, must have formed very close together, and are within 1.2\% of being the same in mass. 

One prediction of a bistability scenario is that magnetic fields should be ubiquitous during the earliest phases of the PMS, when the star is at least partially convective, and should rapidly disappear once stars become radiative. This is precisely what was found by \cite{2019A&A...622A..72V} in their study of intermediate mass T Tauri stars (IMTTS). They showed that essentially all IMTTS with convective envelopes are magnetic, while very few of the fully radiative IMTTS host magnetic fields. \citeauthor{2019A&A...622A..72V} furthermore characterized the magnetic fields of the majority (10/14 or 71\%) of IMTTS with convective envelopes as `complex', i.e.\ possessing significant non-dipolar components, with only one star (7\%) possessing an apparently dipolar field (the remainder could not be classified one way or the other). The frequency of dipolar magnetic fields amongst convective IMTTS is comparable to the incidence of magnetic fields on the main sequence. The identification of a convective boundary for IMTTS magnetic fields is consistent with the results of the study of Herbig Ae/Be stars by \cite{2013MNRAS.429.1001A,2013MNRAS.429.1027A}, who found that the magnetic incidence amongst this population is similar to that of main sequence stars.

It should be noted that the 10\% frequency of fossil fields in hot stars is much lower than the 60\% occurrence of dipoles in M dwarves reported by \cite{2017NatAs...1E.184S}. This suggests either that only the strongest fields survive the transition from convective to radiative regimes, and/or that other factors (such as the ratio of toroidal to poloidal magnetic energy; \citealt{2009MNRAS.397..763B,2010ApJ...724L..34D}) become salient once the stellar magnetic field is no longer supported by a contemporaneous dynamo. 

\section{Conclusions}\label{sec:conclusions}

We report the discovery via KELT photometry of the second chemically peculiar magnetic eclipsing binary system, HD\,62658. Modelling of radial velocities and the TESS light curve reveals that the system is nearly circularized, and that the two components have almost identical masses of about 3 \msun. The out-of-eclipse variability is consistent with rotational modulation by chemical spots, and the evidence suggests that the rotation of the chemically peculiar component is synchronized with the orbit. 

High-resolution spectropolarimtery reveals the system to be an SB2, as expected. One of the components exhibits strong line profile variations consistent with the presence of chemical spots. The other component exhibits weaker variations, which may be consistent with gravity-mode pulsations detected in the light curve. A magnetic field is detected in the chemically peculiar component; \bz~phases coherently with the rotational period inferred from the light curve. Assuming a dipolar oblique rotator model, the magnetic component possesses a surface dipole strength of about 700 G. No magnetic field is detected in the other component, and direct modelling of its circular polarization profile indicates a surface dipole field below about 100 G. 

The existence of two coeval stars with essentially identical fundamental parameters, which formed in the same environment, and that differ only in that one is magnetic, suggests that environmental or evolutionary scenarios for the origin of fossil fields and their rarity in binary systems may be unnecessary, and that the explanation of these phenomena may instead be found in a pre-main sequence dynamo bistability similar to that identified in M-dwarves. It is, however, essential that further observations of this system be obtained, in order to improve the constraints on the magnetic field of the secondary and definitively rule out the presence of a magnetic field on this star. 

This system represents a unique opportunity to compare stellar structure models of stars with and without strong magnetic fields, and may be important for exploration of the consequences of fossil magnetism above and beyond the presence of chemical spots. Such investigations will require detailed knowledge of the magnetic star's surface magnetic field topology and chemical element distribution. 






\section*{Acknowledgements}

This project makes use of data from the KELT survey, including support from The Ohio State University, Vanderbilt University, and Lehigh University, along with the KELT follow-up collaboration. This work has made use of the VALD database, operated at Uppsala University, the Institute of Astronomy RAS in Moscow, and the University of Vienna. This work is based on observations obtained at the Canada-France-Hawaii Telescope (CFHT) which is operated by the National Research Council of Canada, the Institut National des Sciences de l'Univers of the Centre National de la Recherche Scientifique of France, and the University of Hawaii. This research has made use of the SIMBAD database, operated at CDS, Strasbourg, France. Some of the data presented in this paper were obtained from the Mikulski Archive for Space Telescopes (MAST). STScI is operated by the Association of Universities for Research in Astronomy, Inc., under NASA contract NAS5-2655. MES acknowledges support from the Annie Jump Cannon Fellowship, supported by the University of Delaware and endowed by the Mount Cuba Astronomical Observatory. CJ acknowledges funding from the European Research Council (ERC) under the European Union's Horizon 2020 research and innovation programme (grant agreement N$^\circ$670519: MAMSIE), as well as from the Research Foundation Flanders (FWO) under grant agreement G0A2917N (BlackGEM). CJ acknowledges the use of computational resources and services provided by the VSC (Flemish Supercomputer Center), funded by the Research Foundation - Flanders (FWO) and the Flemish Government – department EWI. JLB acknowledges support from FAPESP (grant 2017/23731-1). ADU acknowledges support from the National Science and Engineering Research Council of Canada (NSERC). OK acknowledges support by the Knut and Alice Wallenberg Foundation (project grant ``The New Milky Way''), the Swedish Research Council (project 621-2014-5720), and the Swedish National Space Board (projects 185/14, 137/17). GAW acknowledges support from a Discovery Grant from NSERC. DJJ was supported through the Black Hole Initiative at Harvard University, through a grant (60477) from the John Templeton Foundation and by National Science Foundation award AST-1440254.

\bibliography{bib_dat.bib}{}

\appendix
\section{Posterior Distributions}

\begin{figure}
   \centering
   \includegraphics[width=0.5\textwidth]{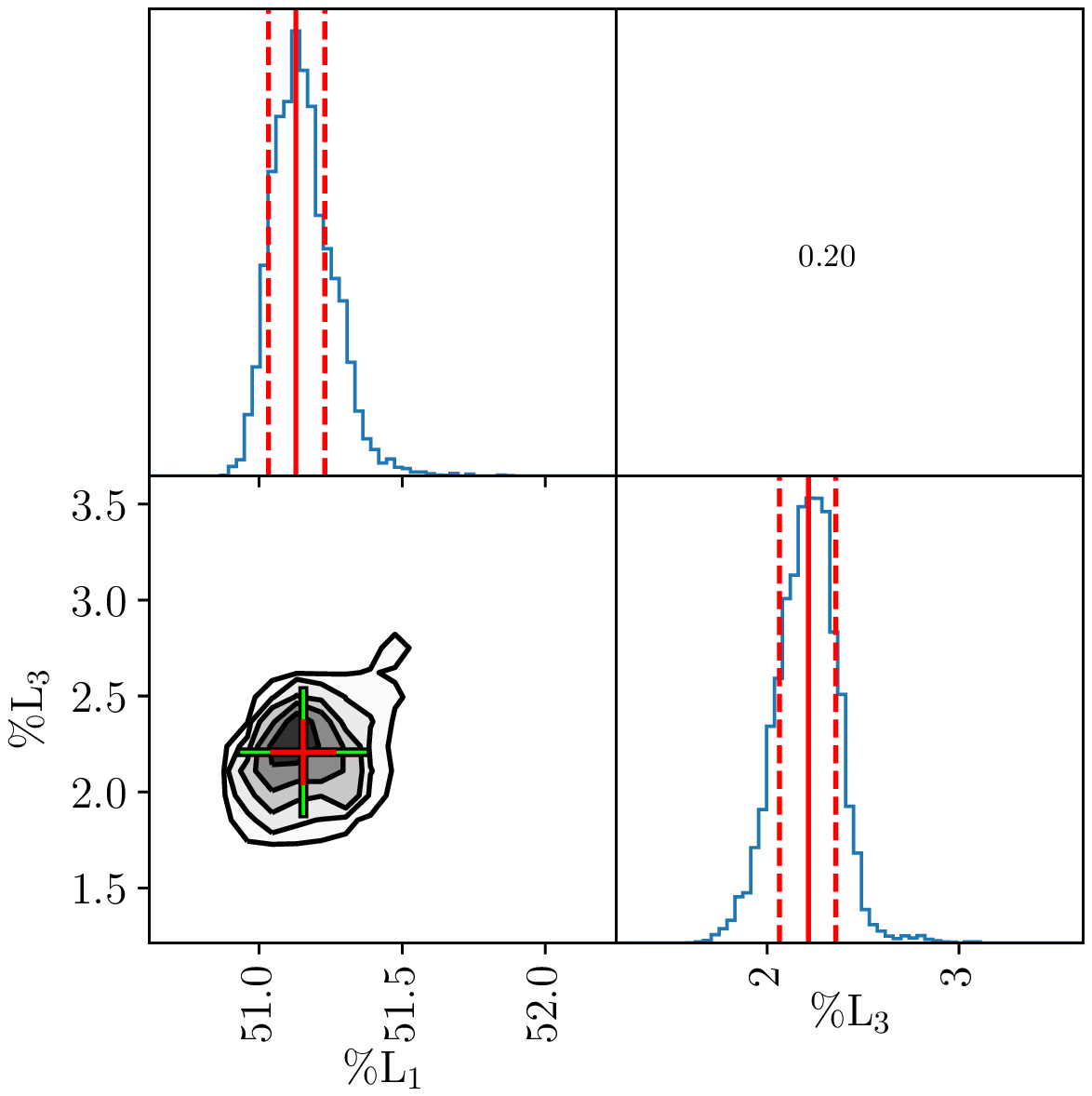}
      \caption[]{Marginalised posterior distributions for the light parameters from MCMC modelling. The number in the top right panel is the correlation coefficient between the corresponding row and column.}
         \label{hd62658_light_posteriors}
   \end{figure}

\begin{figure}
   \centering
   \includegraphics[width=0.5\textwidth]{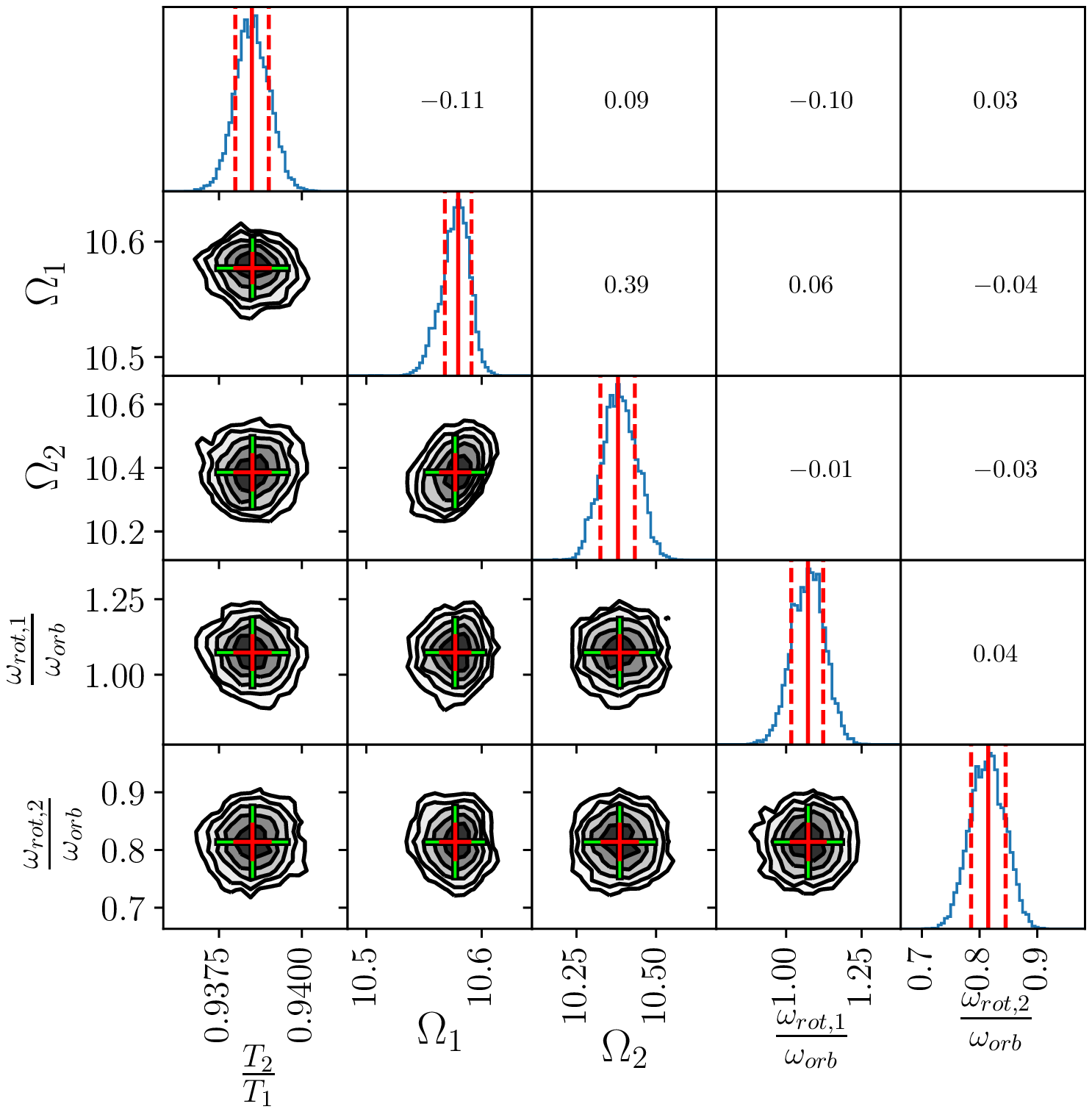}
      \caption[]{Marginalised posterior distributions for the stellar parameters from MCMC modelling.}
         \label{hd62658_star_posteriors}
   \end{figure}

\begin{figure}
   \centering
   \includegraphics[width=0.5\textwidth]{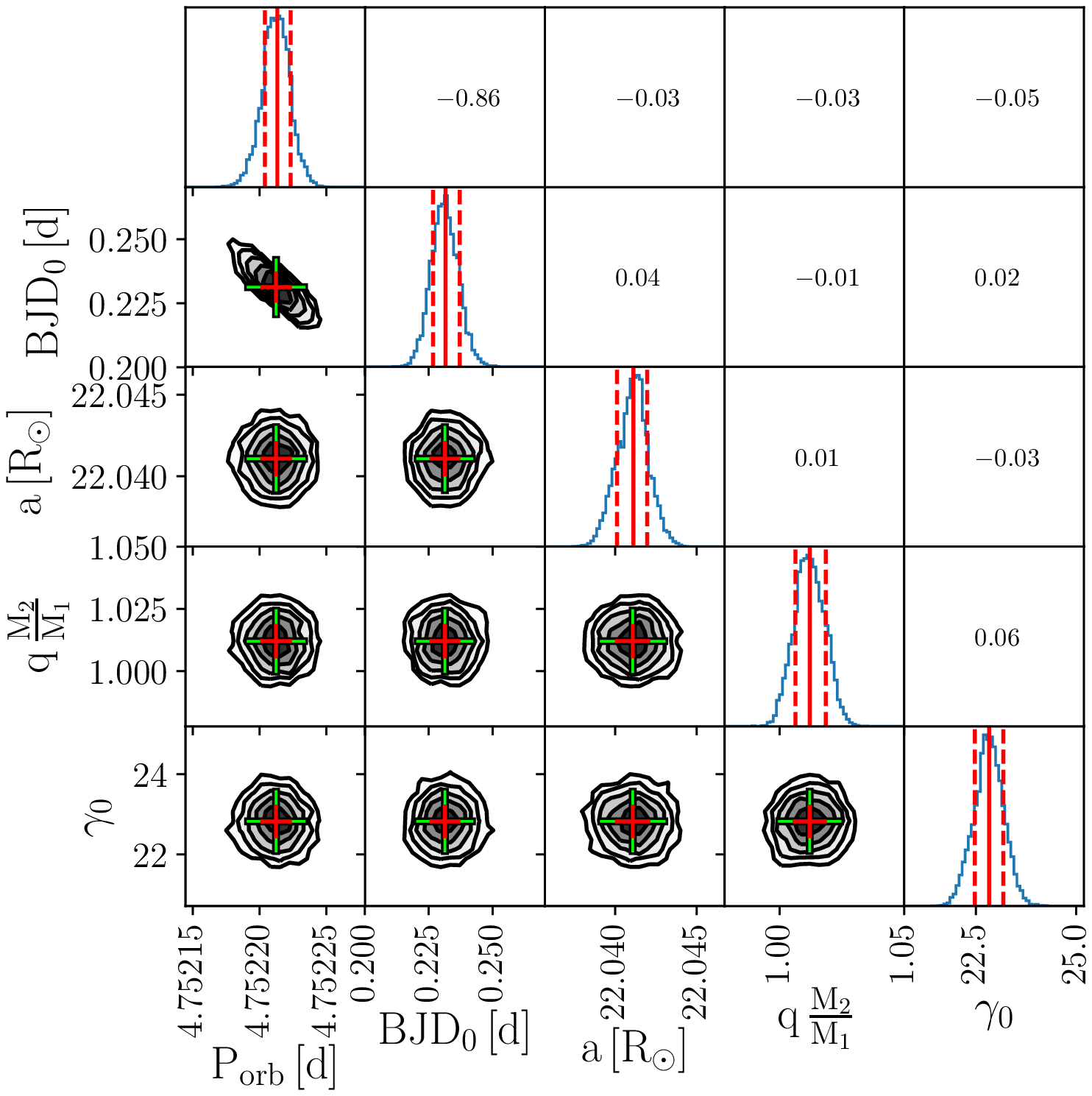}
      \caption[]{Marginalised posterior distributions for the orbital parameters from MCMC modelling.}
         \label{hd62658_orbital_posteriors}
   \end{figure}
   
   \begin{figure}
   \centering
   \includegraphics[width=0.5\textwidth]{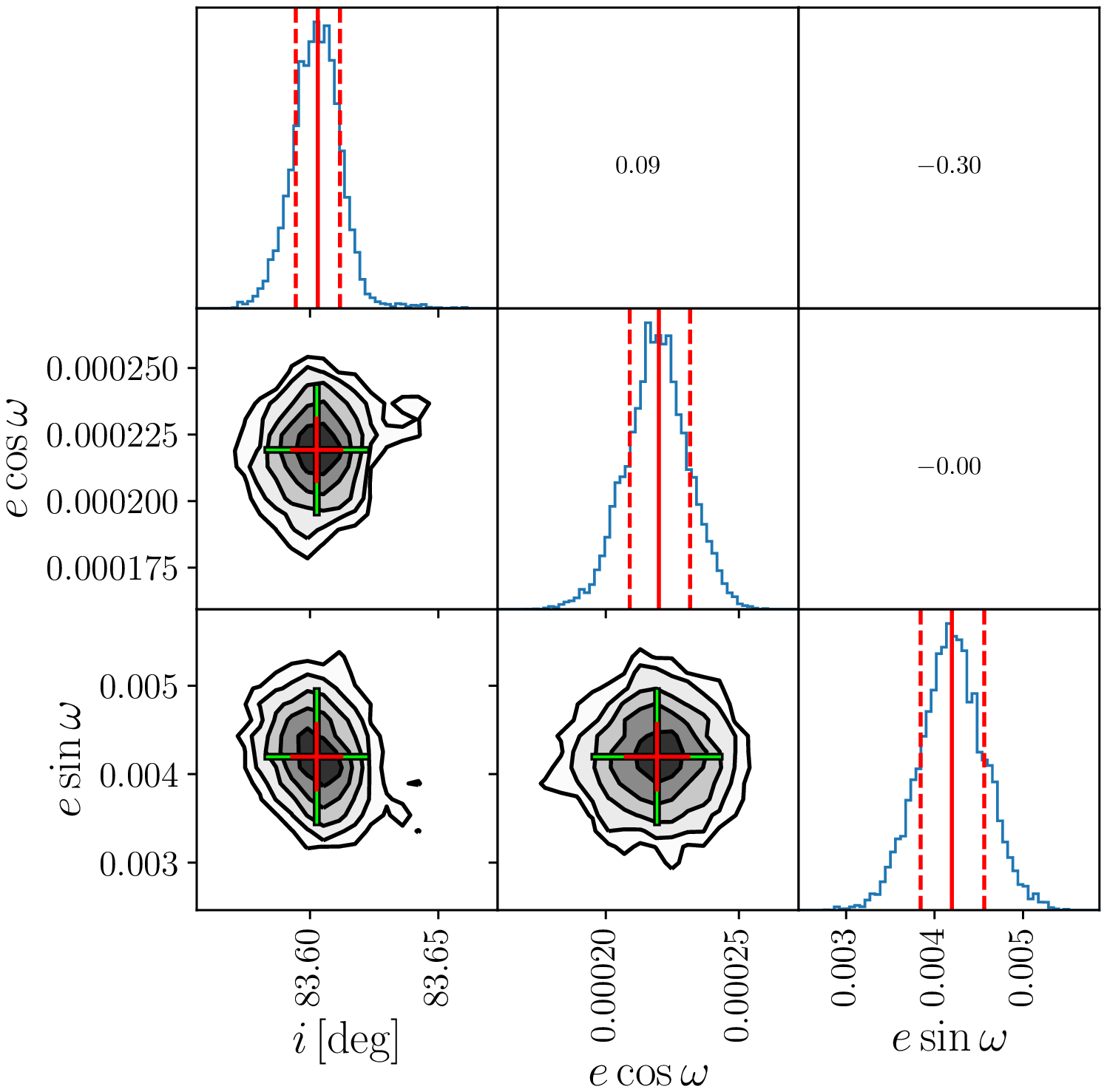}
      \caption[]{Marginalised posterior distributions for more orbital parameters from MCMC modelling.}
         \label{hd62658_orientation_posteriors}
   \end{figure}

\end{document}